\documentclass[12pt]{article}

\usepackage{graphics,amssymb,epsfig,float}
\usepackage[usenames,dvips]{color}
\usepackage{graphicx}
\usepackage{epsfig}
\usepackage{rotating}
\usepackage{dcolumn}
\usepackage{bm}
\usepackage{cite}
\usepackage{amsmath}

\def\lsim{\;\raise0.3ex\hbox{$<$\kern-0.75em\raise-1.1ex\hbox{$\sim$}}\;}
\def\gsim{\;\raise0.3ex\hbox{$>$\kern-0.75em\raise-1.1ex\hbox{$\sim$}}\;}
\def\beq{\begin{equation}}   \def\eeq{\end{equation}}
\def\ba{\begin{array}}       \def\ea{\end{array}}
\def\bea{\begin{eqnarray}}   \def\eea{\end{eqnarray}}
\def\nn{\nonumber}

\begin{document}

\begin{titlepage}
\begin{flushright}
LPT Orsay 12-27 \\ 
LUPM: 12-010
\end{flushright}

\begin{center}
\vspace{1cm}
{\Large\bf Higgs bosons near 125 GeV in the NMSSM with
constraints at the GUT scale} \\
\vspace{2cm}

{\bf{Ulrich Ellwanger$^a$ and Cyril
Hugonie$^b$}}\\
\vspace{1cm}
\it  $^a$ LPT, UMR 8627, CNRS, Universit\'e de Paris--Sud, 91405 Orsay, France \\
\it $^b$ LUPM, UMR 5299, CNRS, Universit\'e de Montpellier II, 34095 Montpellier, France \\
\end{center}

\vspace{1cm}
\begin{abstract}
We study the NMSSM with universal Susy breaking terms (besides the
Higgs sector) at the GUT scale. Within this constrained parameter space,
it is not difficult to find a Higgs boson with a mass of about 125~GeV and
an enhanced cross section in the diphoton channel. An additional lighter
Higgs boson with reduced couplings and a mass $\lsim 123$~GeV is
potentially observable at the LHC. The NMSSM-specific Yukawa couplings
$\lambda$ and $\kappa$ are relatively large and $\tan\beta$ is small, such
that $\lambda$, $\kappa$ and the top Yukawa coupling are of ${\cal O}(1)$
at the GUT scale. The lightest stop can be as light as $105$~GeV, and the
fine-tuning is modest. WMAP constraints can be satisfied by a dominantly
higgsino-like LSP with substantial bino, wino and singlino admixtures and
a mass of $\sim 60-90$~GeV, which would potentially be detectable by
XENON100.
\end{abstract}

\end{titlepage}

\section{Introduction}

Recently, the ATLAS~\cite{:2012si,:2012sk,ATLAS-CONF-2012-019} and
CMS~\cite{Chatrchyan:2012tx,Chatrchyan:2012tw,CMS-PAS-HIG-12-001}
collaborations have presented evidence for a Higgs boson with a mass
near 126~GeV (ATLAS) and 125~GeV (CMS in~\cite{CMS-PAS-HIG-12-001}).
Interestingly, the best fit to the signal strength $\sigma^{\gamma\gamma}
\equiv \sigma_{prod}(H) \times BR(H \to \gamma\,\gamma)$ in the
$\gamma\,\gamma$ search channel is about one standard deviation larger
than expected in the Standard Model (SM);
$\sigma_{obs}^{\gamma\gamma}/\sigma_{SM}^{\gamma\gamma} \sim 2$
for ATLAS~\cite{:2012si,:2012sk}, and
$\sigma_{obs}^{\gamma\gamma}/\sigma_{SM}^{\gamma\gamma} \sim 1.6$
for CMS~\cite{CMS-PAS-HIG-12-001}.

Since then, several publications have studied the impact of a Higgs
boson in the 125~GeV range on the parameter space of supersymmetric
(Susy) extensions of the SM~\cite{
Hall:2011aa,Baer:2011ab,Feng:2011aa,Heinemeyer:2011aa,Arbey:2011ab,
Arbey:2011aa,Draper:2011aa,Moroi:2011aa,Carena:2011aa,Ellwanger:2011aa,
Buchmueller:2011ab,Akula:2011aa,Kadastik:2011aa,Cao:2011sn,
Arvanitaki:2011ck,Gozdz:2012xx,Gunion:2012zd,FileviezPerez:2012iw,
Karagiannakis:2012vk,King:2012is,Kang:2012tn,Chang:2012gp,
Aparicio:2012iw,Roszkowski:2012uf,Ellis:2012aa,Baer:2012uy,Desai:2012qy,
Cao:2012fz,Maiani:2012ij,Cheng:2012np,Christensen:2012ei,Vasquez:2012hn}.
Whereas a Higgs boson in the 125~GeV range is possible within the
parameter space of the Minimal Susy SM (MSSM)~\cite{
Hall:2011aa,Baer:2011ab,Feng:2011aa,Heinemeyer:2011aa,Arbey:2011ab,
Arbey:2011aa,Draper:2011aa,Carena:2011aa,
Buchmueller:2011ab,Akula:2011aa,Kadastik:2011aa,Cao:2011sn,
Arvanitaki:2011ck, Karagiannakis:2012vk,
Aparicio:2012iw,Roszkowski:2012uf,Ellis:2012aa,Baer:2012uy,Desai:2012qy,
Cao:2012fz,Maiani:2012ij,Cheng:2012np,Christensen:2012ei}, large
radiative corrections involving heavy top squarks are required, which
aggravates the ``little fine-tuning problem'' of the MSSM. In addition,
it would be difficult to explain a large enhancement of the diphoton
signal strength in the MSSM~\cite{Carena:2011aa,Cao:2012fz,Christensen:2012ei}.

Within the Next-to-Minimal Supersymmetric SM
(NMSSM~\cite{Maniatis:2009re,Ellwanger:2009dp}),
a Higgs boson in the 125~GeV range is much more
natural~\cite{Hall:2011aa,Ellwanger:2011aa,Arvanitaki:2011ck,King:2012is,Kang:2012tn,Cao:2012fz}:
Additional tree-level contributions and large singlet-doublet mixings
in the CP-even Higgs sector can push up the mass of the mostly
SM-like Higgs boson and, simultaneously, reduce its coupling to
$b$-quarks which results in a substantial enhancement of its
branching fraction into two photons~\cite{Hall:2011aa,Ellwanger:2011aa,King:2012is,Cao:2012fz}.
Studies of the parameter space of the general NMSSM -- including
the dark matter relic density and dark matter nucleon cross section
-- were performed in~\cite{Cao:2012fz,Vasquez:2012hn}.

An important question is whether these interesting features of the
NMSSM survive universality constraints on the soft Susy breaking
parameters at the GUT scale. Since this approach imposes severe
restrictions on the sparticle masses and couplings, it allows to study
whether these would be consistent with present constraints from
direct and indirect sparticle searches, the dark matter relic density
and dark matter direct detection experiments. Moreover it allows to
make predictions for future searches, both in the sparticle and the
Higgs sector. 

The (fully constrained) CNMSSM~\cite{Djouadi:2008yj,Djouadi:2008uj}
was analysed in~\cite{Arbey:2011ab} with the result that, once a relic
density in agreement with WMAP~\cite{Komatsu:2010fb} is imposed,
the Higgs boson mass can barely be above 123~GeV. We find that one
should allow for deviations from full universality in the Higgs sector,
both for the NMSSM-specific soft Susy breaking terms and the
MSSM-like Higgs soft masses like in the MSSM studies
in~\cite{Baer:2011ab,Arbey:2011ab,Buchmueller:2011ab,Cao:2011sn}.
In analogy to the NUHM version of the MSSM,
we shall refer to such a model as NUH-NMSSM for non-universal
Higgs NMSSM. A first study of the NUH-NMSSM was made
in~\cite{Gunion:2012zd} which was confined, however, to the more
MSSM-like region of the parameter space of the NMSSM involving
small values of the NMSSM-specific coupling $\lambda$, and hence
small NMSSM-specific effects in the CP-even Higgs sector.

In the present paper we study the NUH-NMSSM
for large values of the NMSSM-specific coupling $\lambda$ (and low
$\tan\beta$) where the singlet-doublet mixing in the CP-even Higgs
sector is large, and we find that the interesting features of the Higgs
sector of the NMSSM observed
in~\cite{Hall:2011aa,Ellwanger:2011aa,Arvanitaki:2011ck,King:2012is,Kang:2012tn,Cao:2012fz}
can remain present, including constraints from searches for squarks
and gluinos from ATLAS and
CMS~\cite{Chatrchyan:2011zy,Aad:2011ib,CMS-PAS-SUS-12-005},
constraints on the dark matter relic density from
WMAP~\cite{Komatsu:2010fb} and on the dark matter nucleon cross
section from XENON100~\cite{Aprile:2011hi}.
\pagebreak

Our results in the Higgs sector originate essentially from the strong
mixing between all three CP-even Higgs states in the NMSSM: first, a
Higgs boson with a mass in the 125~GeV range can have an
enhanced diphoton signal strength up to
$\sigma_{obs}^{\gamma\gamma}/\sigma_{SM}^{\gamma\gamma} \sim 2.8$.
Second, a lighter less SM-like Higgs boson $H_1$ exists, with small
couplings to electroweak gauge bosons if $M_{H_1} \lsim 114$~GeV
(complying with LEP constraints~\cite{Schael:2006cr}), but a possibly
detectable production cross section at the LHC if $M_{H_1} \gsim
114$~GeV. In fact, a strongly enhanced diphoton signal strength $\gsim
2$ of the Higgs boson $H_2$ with its mass in the 125~GeV range is
possible only if $M_{H_1} \gsim 90$~GeV. The heaviest CP-even Higgs
boson $H_3$, like the heaviest MSSM-like CP-odd and charged Higgs
bosons, have masses in the $250-650$~GeV range, while the lightest
mostly singlet-like CP-odd Higgs state has a mass in the $160-400$~GeV
range. These comply with constraints both from $B$-physics and direct
Susy Higgs searches also for lower masses due to the low values of
$\tan\beta$ considered here and the large singlet component of the
lightest CP-odd state.

In the sparticle sector we require masses for the gluino and the first
generation squarks to comply with constraints from present direct
searches~\cite{Chatrchyan:2011zy,Aad:2011ib,CMS-PAS-SUS-12-005},
but we also study the effect of a reduced sensitivity due to the more
complicated decay cascades in the NMSSM~\cite{Das:2012rr}.
The lightest top squark $\tilde{t}_1$ can be as light as $\sim 105$~GeV
(still satisfying constraints from ATLAS~\cite{ATLAS-stop} and the
Tevatron~\cite{Tevatron-stop}, the latter due to its dominant decay into
a chargino and a b-quark), and the required fine-tuning among the
parameters at the GUT scale remains modest. In the neutralino sector,
the mixings among the five states (bino, wino, two higgsinos and
the singlino) are large. The LSP, with a dominant higgsino component
and a mass of $60-90$~GeV, has a relic density complying with the
WMAP constraints~\cite{Komatsu:2010fb}, and a direct detection
cross section possibly within the reach of XENON100~\cite{Aprile:2011hi}.
However, the supersymmetric contribution to the anomalous magnetic
moment of the muon is somewhat smaller than desired to account for
the deviation of the measurement~\cite{Bennett:2006fi} from the SM.
	
In the next Section we present the analysed parameter space of the
NMSSM with boundary conditions at the GUT scale and the imposed
phenomenological constraints; our results are given in Section~3,
and conclusions in Section~4.

\section{The NMSSM with constraints at the GUT scale}

The NMSSM differs from the MSSM due to the presence of the gauge
singlet superfield $S$. In the simplest $Z_3$ invariant realisation of the
NMSSM, the Higgs mass term $\mu H_u H_d$ in the superpotential
$W_\mathrm{MSSM}$ of the MSSM is replaced by the coupling $\lambda$
of $S$ to $H_u$ and $H_d$ and a self-coupling $\kappa S^3$.  Hence, in
this simplest version the superpotential $W_\mathrm{NMSSM}$ is scale
invariant, and given by
\beq\label{eq:1}
W_\mathrm{NMSSM} = \lambda \hat S \hat H_u\cdot \hat H_d + \frac{\kappa}{3} 
\hat S^3 + \dots\; ,
\eeq
where hatted letters denote superfields, and the ellipsis denote the
MSSM-like Yukawa couplings of $\hat H_u$ and $\hat H_d$ to the
quark and lepton superfields. Once the real scalar component of
$\hat S$ develops a vev $s$, the first term in $W_\mathrm{NMSSM}$
generates an effective $\mu$-term
\beq\label{eq:2}
\mu_\mathrm{eff}=\lambda\, s\; .
\eeq

The soft Susy breaking terms consist of mass terms for the Higgs bosons
$H_u$, $H_d$, $S$, squarks
$\tilde{q_i} \equiv (\tilde{u_i}_L, \tilde{d_i}_L$), $\tilde{u_i}_R^c$,
$\tilde{d_i}_R^c$ and sleptons $\tilde{\ell_i} \equiv (\tilde{\nu_i}_L,
\tilde{e_i}_L$) and $\tilde{e_i}_R^c$ 
(where $i=1..3$ is a generation index):
\bea
-{\cal L}_\mathrm{0} &=&
m_{H_u}^2 | H_u |^2 + m_{H_d}^2 | H_d |^2 + 
m_{S}^2 | S |^2 +m_{\tilde{q_i}}^2|\tilde{q_i}|^2 
+ m_{\tilde{u_i}}^2|\tilde{u_i}_R^c|^2
+m_{\tilde{d_i}}^2|\tilde{d_i}_R^c|^2\nn \\
&& +m_{\tilde{\ell_i}}^2|\tilde{\ell_i}|^2
+m_{\tilde{e_i}}^2|\tilde{e_i}_R^c|^2\; ,
\label{eq:3}
\eea
trilinear interactions involving the third generation squarks, sleptons and the
Higgs fields (neglecting the Yukawa couplings of the two first generations):
\bea
-{\cal L}_\mathrm{3}&=& 
\Bigl( h_t A_t\, Q\cdot H_u \: \tilde{u_3}_R^c +
h_b  A_b\, H_d \cdot Q \: \tilde{d_3}_R^c +
h_\tau A_\tau \,H_d\cdot L \: \tilde{e_3}_R^c  \nn \\
&& +\,  \lambda A_\lambda\, H_u  \cdot H_d \,S +  \frac{1}{3} \kappa 
A_\kappa\,  S^3 \Bigl)+ \, \mathrm{h.c.}\;,
\label{eq:4}
\eea
and mass terms for the gauginos $\tilde{B}$ (bino), $\tilde{W}^a$
(winos) and $\tilde{G}^a$ (gluinos):
 \beq\label{eq:5}
-{\cal L}_\mathrm{1/2}= \frac{1}{2} \bigg[ 
 M_1 \tilde{B}  \tilde{B}
\!+\!M_2 \sum_{a=1}^3 \tilde{W}^a \tilde{W}_a 
\!+\!M_3 \sum_{a=1}^8 \tilde{G}^a \tilde{G}_a   
\bigg]+ \mathrm{h.c.}\;.
\eeq

Expressions for the mass matrices of the physical CP-even and CP-odd
Higgs states -- after $H_u$, $H_d$ and $S$ have assumed vevs $v_u$,
$v_d$ and $s$ and including the dominant radiative corrections -- can be
found in~\cite{Ellwanger:2009dp} and will not be repeated here.
The couplings of the CP-even Higgs states depend on their decompositions
into the weak eigenstates $H_d$, $H_u$ and $S$, which are denoted by
\bea\label{eq:6}
H_1 &=& S_{1,d}\ H_d + S_{1,u}\ H_u +S_{1,s}\ S\; ,\nn \\
H_2 &=& S_{2,d}\ H_d + S_{2,u}\ H_u +S_{2,s}\ S\; ,\nn \\
H_3 &=& S_{3,d}\ H_d + S_{3,u}\ H_u +S_{3,s}\ S\; .
\eea
Then the reduced tree level couplings (relative to a SM-like Higgs
boson) of $H_i$ to $b$ quarks, $\tau$ leptons, $t$ quarks and
electroweak gauge bosons $V$ are
\bea
\frac{g_{H_i bb}}{g_{H_{SM} bb}} = \frac{g_{H_i \tau\tau}}{g_{H_{SM}
\tau\tau}} &=& \frac{S_{i,d}}{\cos\beta}\;,\qquad
\frac{g_{H_i tt}}{g_{H_{SM} tt}} = \frac{S_{i,u}}{\sin\beta}\;,\nn \\
\bar{g}_i \equiv \frac{g_{H_i VV}}{g_{H_{SM} VV}} &=& \cos\beta\,
S_{i,d} + \sin\beta\, S_{i,u}\; .
\label{eq:7}
\eea
Mixings between the SU(2)-doublet and singlet sectors are always
proportional to $\lambda$, and can be sizeable for $\lambda \gsim 0.3$.

As compared to two independent parameters in the Higgs sector of the
MSSM at tree level (often chosen as $\tan \beta$ and $M_A$), the Higgs
sector of the NMSSM is described by the six parameters
\beq \label{eq:8}
\lambda\ , \ \kappa\ , \ A_{\lambda} \ , \ A_{\kappa}, \ \tan\beta\equiv
v_u/v_d\ ,\ \mu_\mathrm{eff}\; .
\eeq
Then the soft Susy breaking mass terms for the Higgs bosons $m_{H_u}^2$,
$m_{H_d}^2$ and $m_{S}^2$ are determined implicitely by $M_Z$,
$\tan\beta$ and $\mu_\mathrm{eff}$.

In constrained versions of the NMSSM (as in the constrained MSSM) one
assumes that the soft Susy breaking terms involving gauginos, squarks or
sleptons are identical at the GUT scale:
\beq \label{eq:9}
M_1 = M_2 = M_3 \equiv M_{1/2}\; ,
\eeq
\beq \label{eq:10}
m_{\tilde{q_i}}^2= m_{\tilde{u_i}}^2=m_{\tilde{d_i}}^2=
m_{\tilde{\ell_i}}^2=m_{\tilde{e_i}}^2\equiv m_0^2\; ,
\eeq
\beq \label{eq:11}
A_t = A_b = A_\tau \equiv A_0\; .
\eeq

In the NUH-NMSSM considered here one allows
the Higgs sector to play a special role: the Higgs soft mass terms
$m_{H_u}^2$, $m_{H_d}^2$ and $m_{S}^2$ are allowed to differ from
$m_0^2$ (and determined implicitely as noted above), and the trilinear
couplings $A_{\lambda}$, $A_{\kappa}$ can differ from $A_0$. Hence the
complete parameter space is characterized by
\beq \label{eq:12}
\lambda\ , \ \kappa\ , \ \tan\beta\ ,\
\mu_\mathrm{eff}\ , \ A_{\lambda} \ , \ A_{\kappa} \ , \ A_0 \ , \ M_{1/2}\ , 
\ m_0\; ,
\eeq
where the latter five parameters are taken at the GUT scale.

Subsequently we are interested in regions of the parameter space
implying large doublet-singlet mixing in the Higgs sector, i.e. large
values of $\lambda$ (and $\kappa$) and low values of $\tan\beta$, which
lead naturally to a SM-like Higgs boson $H_2$ in the 125~GeV range~\cite{
Hall:2011aa,Ellwanger:2011aa,Arvanitaki:2011ck,King:2012is,Kang:2012tn,
Cao:2012fz}. Requiring $124\ \mathrm{GeV} < M_{H_2} < 127\ \mathrm{GeV}$
and  $\sigma_{obs}^{\gamma\gamma}(H_2)/\sigma_{SM}^{\gamma\gamma} > 1$,
we find
\bea
0.41 \, < & \lambda & < \, 0.69\; ,\nn \\
0.21 \, < & \kappa & < \, 0.46\; ,\nn \\
1.7 \, < & \!\!\!\!\tan\beta\!\!\!\! & < \, 6\;
\label{eq:13}
\eea
(with many points for $\tan\beta \lsim 2.5$).
It is intriguing that with these choices at the weak scale, one obtains
$\lambda \sim \kappa \sim h_t \sim {\cal O}(1)$ for the running
couplings at the GUT scale; hence all 3 Yukawa couplings are close to
(but still below) a Landau singularity.

We assume $\mu_\mathrm{eff} >0$ which implies $A_0,\ A_\lambda,\
A_\kappa < 0$.
Constraints on the soft Susy breaking parameters depend strongly on the
sparticle decay cascades. Using the absence of signal at the LHC in the jets and
missing transverse momentum search channels, bounds in the $m_0, M_{1/2}$
plane have been derived in the CMSSM with $\tan\beta =
10$~\cite{Chatrchyan:2011zy,Aad:2011ib,CMS-PAS-SUS-12-005}.
In the NUH-NMSSM, however, we find lighter stops (due to the
lower values of $\tan\beta$ implying a larger value of the top Yukawa
coupling, which affects the RGE running of the soft Susy breaking stop
masses), and a modified neutralino sector which reduces the sensitivity
in these search channels \cite{Das:2012rr}.

Hence, to start with, we impose only
constraints from sparticle searches at LEP~\cite{LEPSUSY} and the
Tevatron~\cite{Abazov:2006wb,Wang:2009zzf}, and from stop searches
at the Tevatron \cite{Tevatron-stop} and the LHC \cite{ATLAS-stop}.
These imply
\bea
m_0 \gsim 140\ \text{GeV}\; , && 
M_{1/2} \gsim 270\ \text{GeV} \nn \\
m_{\tilde{q}} \gsim 580\ \text{GeV}\; , &&
M_{\tilde{g}} \;\; \gsim 640\ \text{GeV}\; .
\label{eq:14}
\eea
In addition we require that the fine-tuning $\Delta$ defined in
eq.~(\ref{eq:17}) satisfies $\Delta<120$, which implies upper bounds
which will be discussed below.
However, it is possible that the stronger CMSSM-like constraints
in the $m_0, M_{1/2}$ plane~\cite{CMS-PAS-SUS-12-005} also
apply to the NUH-NMSSM considered here. These stronger
constraints further reduce the allowed points in the parameter
space approximately to
\beq
m_{\tilde{q}} \gsim 1250\ \text{GeV}\; , \quad
M_{\tilde{g}} \gsim 850\ \text{GeV}\; ,
\label{eq:15}
\eeq
but we scanned the constraints from~\cite{CMS-PAS-SUS-12-005} in the
$m_0, M_{1/2}$ plane exactly. These constraints are used for the points
shown in the Figs.~(\ref{fig:1}-\ref{fig:3}) below, but
the difference between the constraints (\ref{eq:14}) and (\ref{eq:15})
has practically no impact on our results in the Higgs sector.
Remarkably, regardless of the constraints in the $m_0, M_{1/2}$ plane,
the lightest stop mass can be as low as $\sim 105$~GeV.

Together with these bounds on $m_0, M_{1/2}$, the above constraints
on the Higgs sector and the LEP bound on the chargino mass lead to
\bea
105 <& \!\!\mu_\mathrm{eff}\!\! &< 205\ \text{GeV}\ \text{for weak constraints}\
(\ref{eq:14})\; ,\nn \\
105 <& \!\!\mu_\mathrm{eff}\!\! &< 160\ \text{GeV}\ \text{for strong constraints}\
(\ref{eq:15})\; .
\label{eq:16}
\eea

We have scanned the parameter space of the NUH-NMSSM given
in~(\ref{eq:12}) using a Markov Chain Monte Carlo (MCMC) technique,
which yields a very large number of points ($\sim 10^6$) satisfying all
the phenomenological constraints described below. To do so, we have
modified the code \hbox{NMSPEC}~\cite{Ellwanger:2006rn} inside
NMSSMTools~\cite{Ellwanger:2004xm,Ellwanger:2005dv} in order
to allow for $\kappa$ and $\mu_\mathrm{eff}$ to be used as input
parameters at the weak scale and to compute the Higgs soft masses
$m_{H_u}^2$, $m_{H_d}^2$ at the GUT scale; a corresponding
version 3.2.0 will be made public soon. In \hbox{NMSPEC}, the two
loop renormalization group equations (RGEs) between the weak
and GUT scales are integrated numerically for all parameters. In
the presence of boundary conditions both at the weak and the GUT
scales as it is the case here, these can be satisfied only through an
iterative process. This iterative process is not guaranteed to converge,
notably for large Yukawa couplings as in~(\ref{eq:13}). In fact, the RGE
integration algorithm within the latest public version 3.1.0 of NMSSMTools
had to be modified in version 3.2.0 to achieve convergence for large Yukawa
couplings.

In the Higgs sector we have used two-loop radiative corrections
from~\cite{Degrassi:2009yq}, and for the top quark pole mass we use
$m_\text{top}=172.9$~GeV.
Our results in the next Section use the reduced Higgs production rates
(normalized with respect to the SM production rates) in various channels.
For gluon-gluon fusion we use the reduced Higgs-gluon coupling as
computed in NMSSMTools, which takes care of all colored (s)particles
in the loop. For the low values of $\tan\beta$ considered here, the top
quark loop dominates by far, and leads essentially to 
${g_{H_itt}}/{g_{H_{SM} tt}}$ as given in~(\ref{eq:7}). Since a single
particle loop dominates, radiative corrections not considered in
NMSSMTools tend to cancel in the ratio to the SM. Likewise, Higgs
production rates via associate production with $W/Z$ ($\equiv V$) or
vector boson fusion (VBF) are simply proportional to the SM rates
rescaled by $\bar{g}_i^2$ defined in~(\ref{eq:7}). The Higgs branching
fractions are computed in NMSSMTools to the same accuracy both
for the NMSSM and a SM-like Higgs boson, such that radiative
corrections not considered in NMSSMTools tend again to cancel in
the ratio to the SM.

Next we turn to the imposed phenomenological constraints. In the Higgs
sector we impose constraints from LEP~\cite{Schael:2006cr}, which still
allow for a Higgs mass below 114~GeV if its coupling to the $Z$~boson
is reduced. Constraints on Higgs bosons from the LHC are those
implemented in the version 3.1.0 of NMSSMTools, which are based on
public ATLAS and CMS results available at the end of 2011, including
constraints from CMS on heavy MSSM-like Higgs bosons decaying to tau
pairs~\cite{Chatrchyan:2012vp}. In version 3.2.0 however, we have updated
the important $\gamma\,\gamma$ search channel with the results from
ATLAS~\cite{:2012sk} and CMS~\cite{Chatrchyan:2012tw}.
In order to fit the evidence of both experiments in the $\gamma\,\gamma$
channel, we impose $124\ \mathrm{GeV} < M_{H} < 127\ \mathrm{GeV}$,
which is satisfied exclusively by $H_2$ for larger values of $\lambda$, 
and we require a good visibility of $H_2$ in the $\gamma\,\gamma$ channel, i.e. 
$\sigma_{obs}^{\gamma\gamma}(H_2)/\sigma_{SM}^{\gamma\gamma} > 1$ in
both the gluon fusion and VBF production modes.

Also the constraints from $B$-physics are those implemented in the
version 3.1.0 of NMSSMTools. In spite of charged (resp.~CP-odd) Higgs
masses as low as $\sim$~250 (resp. 160)~GeV, these are easily satisfied
for low values of $\tan\beta$, or a large singlet component for the lightest
CP-odd state (i.e. the couplings of these Higgs bosons to
$b$-quarks are hardly enhanced with respect to the SM Higgs).

The dark matter relic density and direct detection cross section of the
LSP $\chi^0_1$ (the lightest neutralino) are computed with the help of
MicrOmegas~\cite{Belanger:2005kh,Belanger:2006is,Belanger:2008sj}
implemented in NMSSMTools. The default constraints $0.094 <
\Omega h^2 < 0.136$ are slightly weaker than the most recent ones from
WMAP~\cite{Komatsu:2010fb}, but this has no impact on the viable regions
in parameter space (only on the number of points retained). We also apply
the bounds from XENON100~\cite{Aprile:2011hi} on the spin
independent $\chi^0_1$-nucleon cross section (roughly $\sigma^{si}(p)
\lsim 10^{-8}$~pb for $M_{\chi^0_1} \sim 60-90$~GeV).

Since we hardly find light sleptons of the second generation (and
again due to the low values of $\tan\beta$), the Susy contribution
$\Delta a_\mu$ to the anomalous magnetic moment of the muon
is below $\Delta a_\mu \lsim 7\cdot 10^{-10}$, violating the constraint
implemented in NMSSMTools. However, it still improves the discrepancy
between the SM and the measured value~\cite{Bennett:2006fi}, and can
reduce the discrepancy to two standard deviations depending on the
employed SM value.

\section{Results}

Some remarks on our results have already been made above. Notably
we have no difficulties to find points in the parameter space satisfying
the above constraints, including the dark matter relic density and the
direct detection cross section, $124\ \mathrm{GeV} < M_{H_2} < 127\
\mathrm{GeV}$ and $R_2^{\gamma\gamma}> 1$, where we define
\beq
R_2^{\gamma\gamma} \equiv
\sigma_{obs}^{\gamma\gamma}(H_2) / \sigma_{SM}^{\gamma\gamma} \; .
\eeq

The mechanism behind this enhancement has been discussed earlier
in~\cite{Hall:2011aa,Ellwanger:2011aa,Ellwanger:2010nf}: the
$BR(H_2 \to \gamma\,\gamma)$ is strongly enhanced due to a reduced
total width (dominated by \linebreak
$\Gamma(H_2 \to b\,\bar{b})$) for a small reduced coupling
${g_{H_2 bb}}/{g_{H_{SM} bb}}$ in~(\ref{eq:7}), i.e. a small value of
the mixing angle $S_{2,d}$, in spite of a milder reduction of the
partial width $\Gamma(H_2 \to \gamma\,\gamma)$. This occurs for large
singlet-doublet mixing (which also leads to an increase of $M_{H_2}$),
and requires that the third eigenstate $H_3$ is not decoupled, i.e. not
too heavy. The enhancement of the $BR(H_2 \to \gamma\,\gamma)$ also
over-compensates a milder reduction of the production cross section of
$H_2$ due to singlet-doublet mixing. 

The reduction of the total width leads also to a potential increase of
the reduced signal rate in the $ZZ/WW$ channels (via gluon fusion)
\beq
R_2^{VV}(gg) \equiv \sigma_{obs}^{ZZ}(gg \to H_2)/\sigma_{SM}^{ZZ}(gg
\to H) \; =\sigma_{obs}^{WW}(gg \to H_2)/\sigma_{SM}^{WW}(gg \to H) \; ,
\eeq
in spite of the reduction of the partial widths $\Gamma(H_2 \to ZZ/WW)$
due to singlet-doublet mixing. A fourth interesting reduced signal cross
section is the $\tau\tau$ channel via VBF 
\beq
R_2^{\tau\tau}(\text{VBF}) \equiv
\sigma_{obs}^{\tau\tau}(WW \to H_2)/
\sigma_{SM}^{\tau\tau}(WW \to H) \; ,
\eeq
which tends to be reduced, however, for a small mixing angle $S_{2,d}$.

\begin{figure}[ht!]
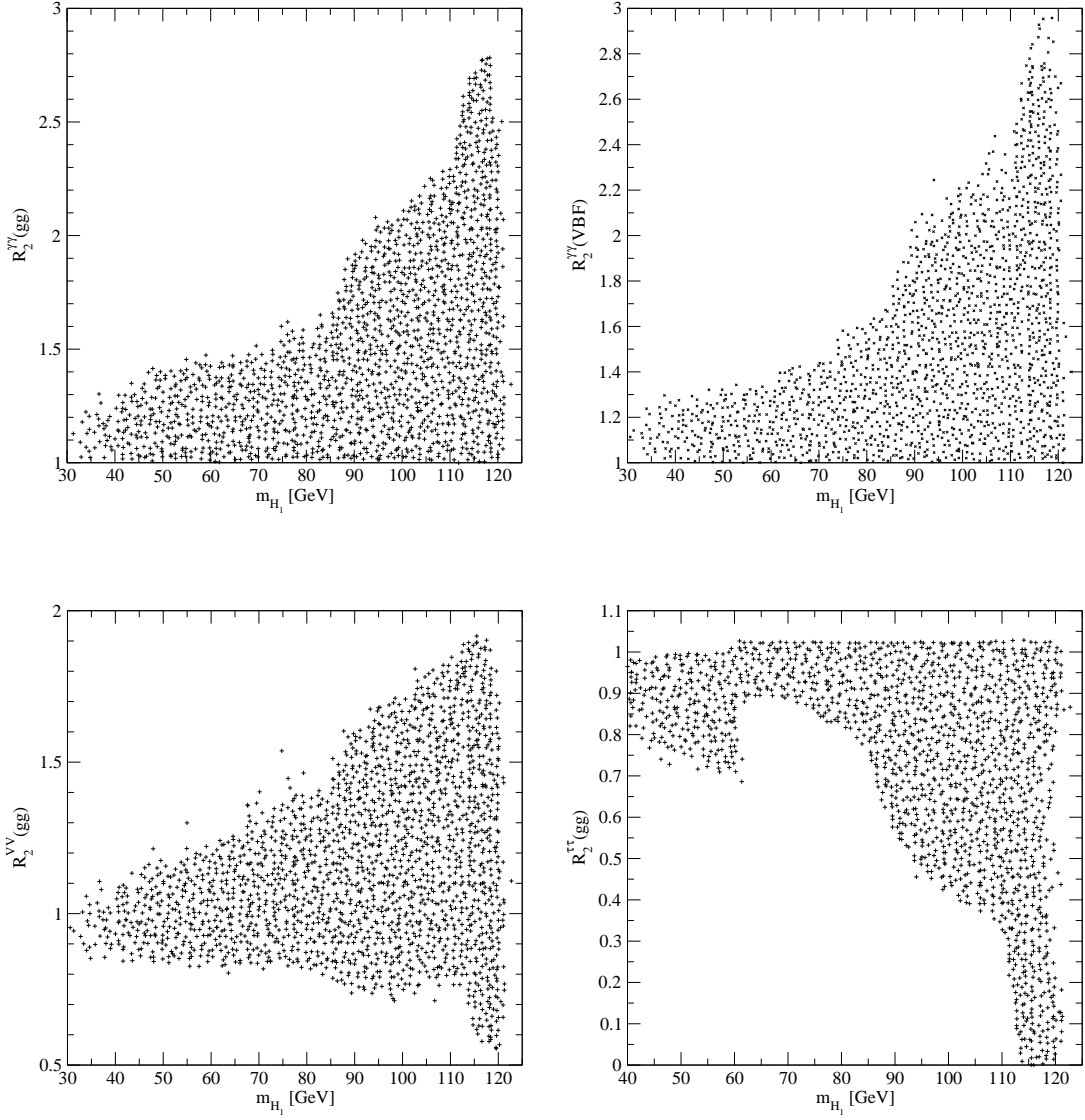

\vspace*{10mm}
\begin{center}
\begin{tabular}{cc}
\psfig{file=fig1a.eps, scale=0.4}
\   &
\psfig{file=fig1b.eps, scale=0.4}
\\ \\  \\ 
\psfig{file=fig1c.eps, scale=0.4}
\   &
\psfig{file=fig1d.eps, scale=0.4}
\end{tabular}
\vspace*{3mm}
\caption{Reduced signal cross sections $R_2$ for $H_2$ with a mass in
the $124-127$~GeV range, as a function of $M_{H_1}$
for a representative sample of viable points in parameter space.
Upper left: $R_2^{\gamma\gamma}(gg)$ (diphoton channel, $H_2$
production via gluon fusion), upper right: $R_2^{\gamma\gamma}(\text{VBF})$
(diphoton channel, $H_2$ production via VBF), lower left: $R_2^{VV}(gg)$
($ZZ, WW$ channels, $H_2$ production via gluon fusion), lower right:
$R_2^{\tau\tau}(\text{VBF})$ ($\tau\,\tau$ channel, $H_2$ production via
VBF).}
\label{fig:1}
\end{center}
\end{figure}

Singlet/doublet mixing angles are typically large, if the eigenstates are
close in mass. Hence we should expect that $R_2^{\gamma\gamma}$ is
the larger, the closer $M_{H_1}$ is to $M_{H_2}$, i.e. the heavier is $H_1$.
Subsequently we consider separately $R_2^{\gamma\gamma}(gg)$ (where
$H_2$ is produced via gluon fusion), and $R_2^{\gamma\gamma}(\text{VBF})$
(where $H_2$ is produced via VBF). In Figs.~\ref{fig:1} we show
$R_2^{\gamma\gamma}(gg)$, $R_2^{\gamma\gamma}(\text{VBF})$,
$R_2^{VV}(gg)$ and $R_2^{\tau\tau}(\text{VBF})$ as a function of $M_{H_1}$ for 
a representative sample of $\sim 2000$ points in the scanned parameter space
of the semi-constrained NMSSM described above. All points satisfy the WMAP
bound on the dark matter relic density, the XENON100 bound on $\sigma^{si}(p)$,
and the stronger lower bound on $M_{1/2}$ given in eq.~(\ref{eq:15}). (Relaxing
this bound to the one given in eq.~(\ref{eq:14}) does not lead to additional regions
in Figs.~\ref{fig:1}.)

We see that, as expected, the ratios $R_2^{\gamma\gamma}(gg)$,
$R_2^{\gamma\gamma}(\text{VBF})$ and $R_2^{VV}(gg)$ can increase with
$M_{H_1}$, and  $R_2^{\gamma\gamma}(gg)$ can become as large as 2.8
for $M_{H_1} \gsim 115$~GeV. (The inverse conclusion does not hold:
$M_{H_1} \gsim 115$~GeV does not imply $R_2^{\gamma\gamma}>2$.)
$R_2^{\tau\tau}(\text{VBF})$ is below $\sim 1$, and the very small
values of $R_2^{\tau\tau}(\text{VBF})$ correspond to the highest values
of $R_2^{\gamma\gamma}(gg)$. For $M_{H_2}$ in the range $124-127$~GeV,
none of these reduced signal rates shows a significant dependency on
$M_{H_2}$; corresponding plots would only transcribe the LHC constraints
on each rate as a function $M_{H_2}$, but they would not provide additional
informations and are hence omitted.

\begin{figure}[ht!]
\vspace{15mm}
\begin{center}
\psfig{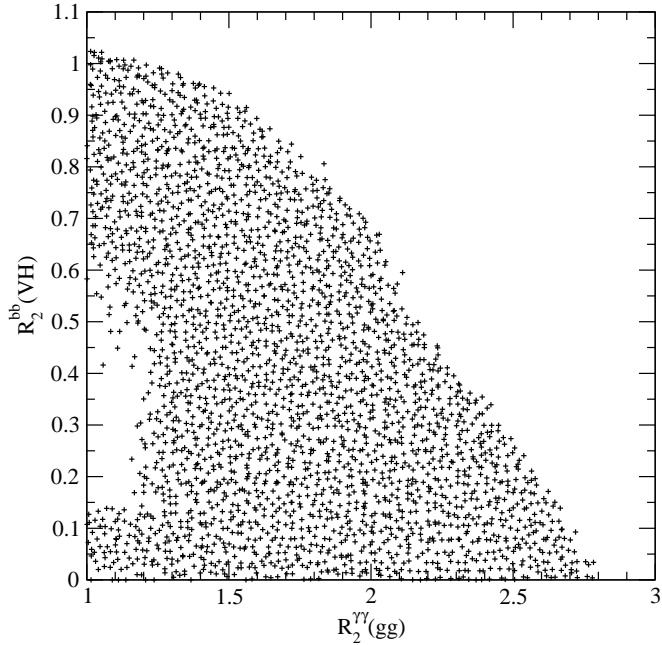}
\vspace*{3mm}
\caption{$R_2^{b\bar{b}}(\text{VH})$ (associate production $W/Z+H_2$
with $H_2\to b\bar{b}$) as a function of $R_2^{\gamma\gamma}(gg)$.}
\label{fig:2}
\end{center}
\end{figure}

Recently, excesses compatible with a Higgs boson in the 125~GeV range
have also been observed at the Tevatron~\cite{Tevatron-higgs}. Here, the
dominant excess originates from associated $VH$ production
with $H \to b \bar{b}$. The corresponding reduced signal rate for the
candidate $H_2$, $R_2^{b\bar{b}}(\text{VH})$ (which is equal to
$R_2^{\tau\tau}(\text{VBF})$), cannot be very small given the observations
at the Tevatron. In Fig.~\ref{fig:2} we show  $R_2^{b\bar{b}}(\text{VH})$
against $R_2^{\gamma\gamma}(gg)$. We see that $R_2^{b\bar{b}}(\text{VH})
\gsim 0.7$ is possible only for $R_2^{\gamma\gamma}(gg)\lsim 2$, but
$R_2^{\gamma\gamma}(gg)\gsim 1.6$ still allows for $R_2^{b\bar{b}}(\text{VH})
\gsim 0.9$.

\begin{figure}[ht!]
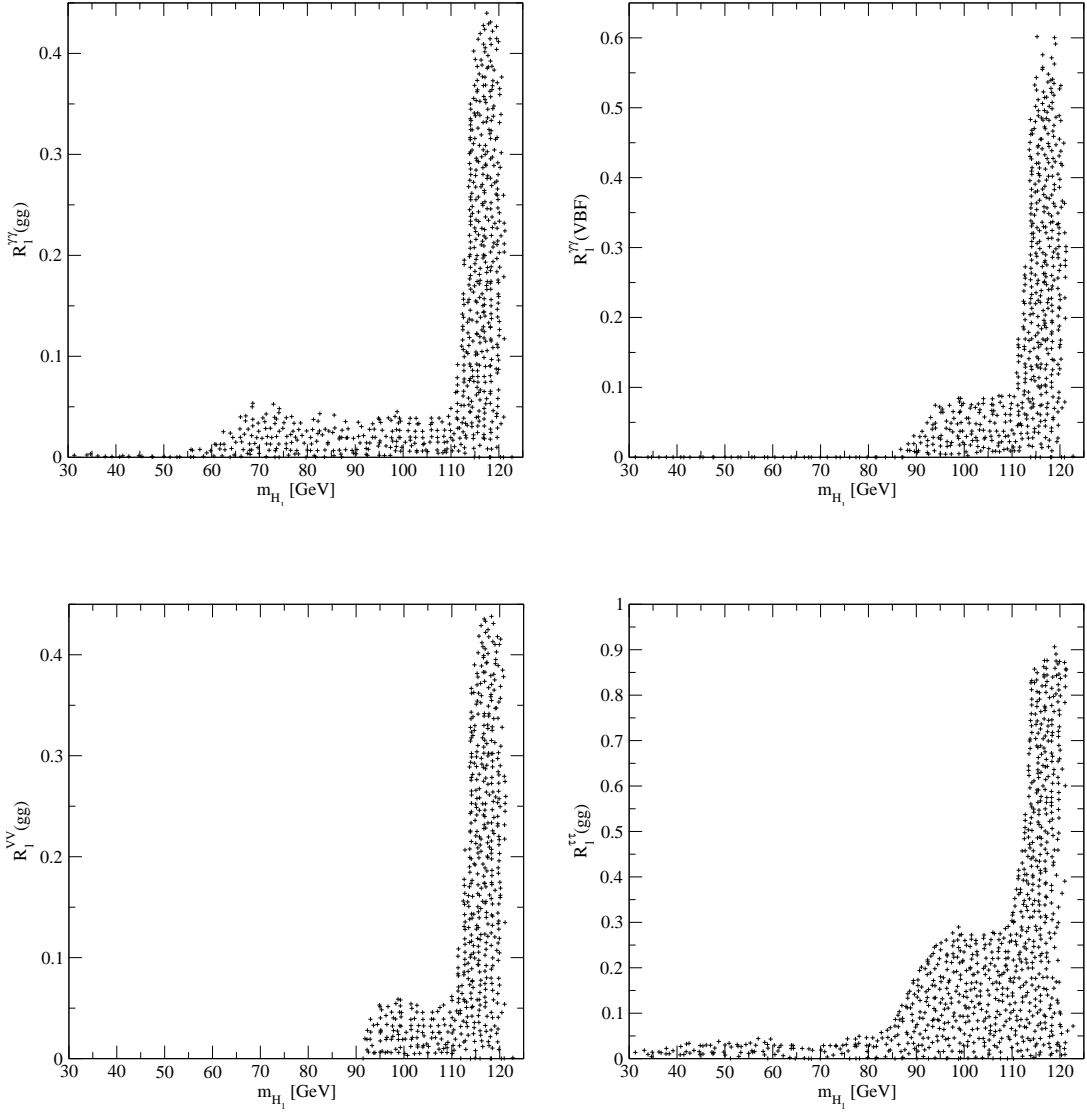

\begin{center}
\begin{tabular}{cc}
\psfig{file=fig2a.eps, scale=0.4}
\   &
\psfig{file=fig2b.eps, scale=0.4}
\\ \\ \\
\psfig{file=fig2c.eps, scale=0.4}
\   &
\psfig{file=fig2d.eps, scale=0.4}
\end{tabular}
\vspace*{3mm}
\caption{Reduced signal cross sections $R_1$ as a function of $M_{H_1}$.
Upper left: $R_1^{\gamma\gamma}(gg)$ (diphoton channel, $H_1$ production
via gluon fusion), upper right: $R_1^{\gamma\gamma}(\text{VBF})$ (diphoton
channel, $H_1$ production via VBF), lower left: $R_1^{VV}(gg)$ ($ZZ$ and
$WW$ channels, $H_1$ production via gluon fusion), lower right:
$R_1^{\tau\tau}(\text{VBF})$ ($\tau\,\tau$ channel, $H_1$ production via VBF).}
\label{fig:3}
\end{center}
\end{figure}

Obviously the reduced signal rates of $H_1$ are also very important. For
instance, $H_1$ could be compatible with the excess of events observed
by CMS for $M_H \sim 119.5$~GeV in the $ZZ$ channel~\cite{Chatrchyan:2012tx}.
On the other hand, for $M_{H} \sim 95-100$~GeV the upper bounds from LEP on
its reduced coupling to the $Z$~boson are particularly weak, and a mostly (but not
completely) singlet-like $H_1$ could explain the mild excess of events observed
there~\cite{Dermisek:2007ah,Ellwanger:2011sk,Schael:2006cr}. The corresponding
reduced signal cross sections as a function of $M_{H_1}$ are shown in Figs.~\ref{fig:3}.
As explained in~\cite{Ellwanger:2011aa}, the reduced signal cross section in the
$b\bar{b}$ channel at LEP coincides with $R^{\tau\tau}(\text{VBF})$.

We see that the reduced signal cross sections are mostly small for
$M_{H_1} \lsim 110$~GeV where the singlet component of $H_1$ is large,
but $R_1^{\tau\tau}(\text{VBF})$ can be as large as $\sim 0.25$ for
$M_{H_1} \sim 95-100$~GeV which is interesting given the mild excess of
events observed at LEP. On the other hand, for $M_{H_1} \gsim 110$~GeV,
$R_1^{\gamma\gamma}(gg)$, $R_1^{\gamma\gamma}(\text{VBF})$ and
$R_1^{VV}(gg)$ can become as large as $\sim 0.5$ and
$R_1^{\tau\tau}(\text{VBF})$ as large as $\sim 0.9$, hence $H_1$ is
potentially detectable.

The Higgs sector of the NMSSM contains a third CP-even state $H_3$,
two CP-odd states $A_1$ and $A_2$ and, as in the MSSM, a charged
Higgs boson $H_\pm$. We find that the lightest CP-odd state $A_1$ is
mostly singlet-like with a mass in the range  $160-400$~GeV, and hardly
visible at the LHC due to its small production cross sections. The states
$H_3$, $A_2$ and $H_\pm$ all have similar masses in the
$250-650$~GeV range and would also be difficult to see at the LHC due
to the small value of $\tan\beta$ in the region of the parameter space of
interest~(\ref{eq:13}).

\begin{figure}[ht!]
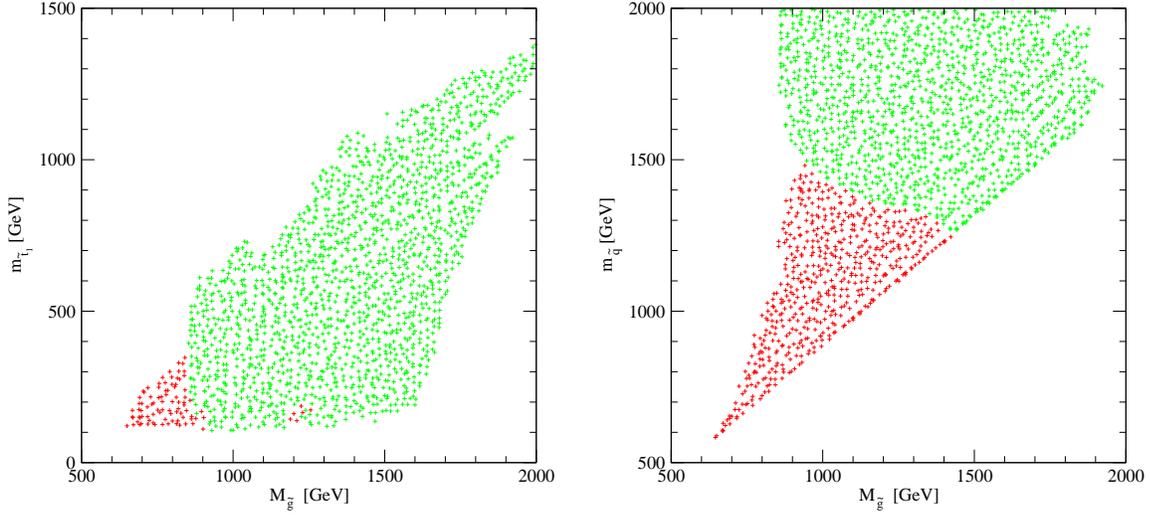

\vspace{10mm}
\begin{center}
\begin{tabular}{cc}
\psfig{file=fig3.eps, scale=0.4}
\   &
\psfig{file=fig3c.eps, scale=0.4,clip=}
\end{tabular}
\vspace*{8mm}
\caption{$m_{\tilde{t}_1}$ (left panel) and $m_{\tilde{q}}$ (right panel)
as a function of $M_{\tilde{g}}$.
Points in red (darker points) satisfy the weaker bounds~(\ref{eq:14}),
but not the stronger bounds~(\ref{eq:15}), while points in green
(brighter) satisfy both constraints.}
\label{fig:4}
\end{center}
\end{figure}

The masses of the sparticles are essentially determined by $M_{1/2}$,
$m_0$, $A_0$ and $\mu_\text{eff}$. In Figs.~\ref{fig:4} we show the mass
$m_{\tilde{q}}$ of the
lightest first generation squark ($d_R$ for our choice
of parameters)
as well as the mass $m_{\tilde{t}_1}$ of the lightest (mostly right-handed)
stop as a function of the gluino mass $M_{\tilde{g}}$. Here it makes a
difference whether we impose the weaker bounds~(\ref{eq:14}) or the
stronger bounds~(\ref{eq:15}):
points in red satisfy only the weaker bounds
while points in green satisfy both constraints.
(For $M_{\tilde{g}}\gsim 640$~GeV, a stop
mass as small as 105~GeV is not excluded by present searches at the
LHC~\cite{ATLAS-stop}, but could become observable in the near future.)

\begin{figure}[ht!]
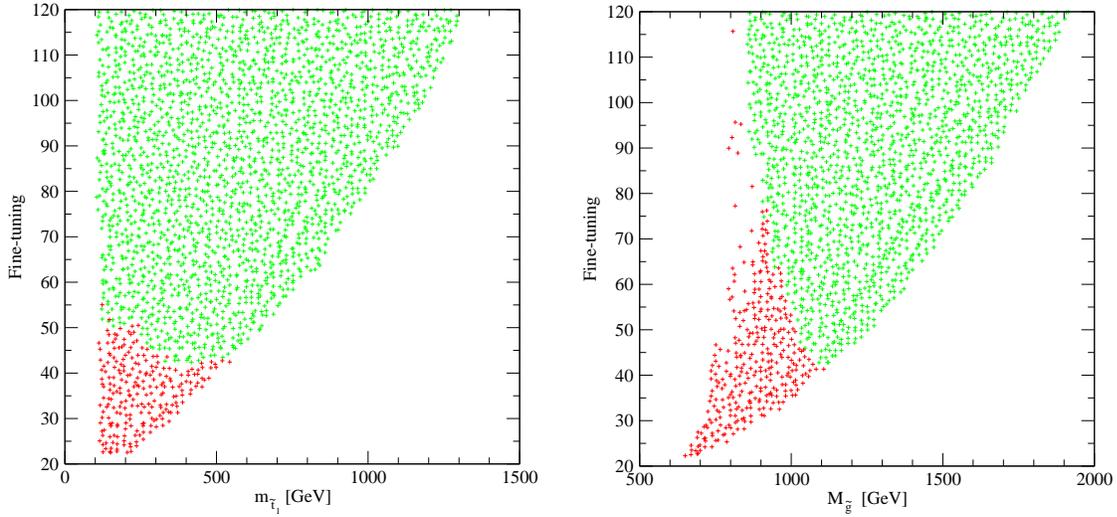

\vspace*{5mm}
\begin{center}
\begin{tabular}{cc}
\psfig{file=fig3b.eps, scale=0.4}
\   &
\psfig{file=fig3a.eps, scale=0.4,clip=}
\end{tabular}
\vspace*{8mm}
\caption{Fine-tuning as a function of $m_{\tilde{t}_1}$ (left panel) and
$M_{\tilde{g}}$ (right panel). The color code is as in Figs.~\ref{fig:4}.}
\vspace*{5mm}
\label{fig:5}
\end{center}
\end{figure}

It is known that the stop mass has an impact on the fine-tuning with
respect to the fundamental parameters of Susy extensions of the SM,
due to its impact on the running soft Susy breaking Higgs mass terms.
In addition, both are affected by the gluino mass. We have
estimated the quantitative amount of fine-tuning with respect to the
parameters at the GUT scale following the procedure outlined
in~\cite{Ellwanger:2011mu}. There, a fine-tuning measure
\beq\label{eq:17}
\Delta = \mathrm{Max}\{\Delta_i^{\mathrm{GUT}}\},\qquad 
\Delta_i^{\mathrm{GUT}} = \left|\frac{\partial \ln(M_Z)}
{\partial \ln(p_i^{\mathrm{GUT}})}\right|
\eeq
was used, where $p_i^{\mathrm{GUT}}$ are 
all parameters at the GUT scale (Yukawa couplings and soft
Susy breaking terms). (Note that sometimes $\ln(M_Z^2)$
instead of $\ln(M_Z)$ is used in the definition of $\Delta$, leading to
an obvious factor of 2). 
We find that $\Delta$, shown as a function of $m_{\tilde{t}_1}$ and
$M_{\tilde{g}}$ in Figs.~\ref{fig:5}, is dominated as usual by
$p_i^{\mathrm{GUT}}=M_{1/2}$.

We see that the fine-tuning $1/\Delta$ can be as low as ${\cal O}(5\%)$
(with the definition in~(\ref{eq:17})) in the range of smaller stop
and gluino masses allowed by the weaker lower bounds~(\ref{eq:14}), and
still as low as ${\cal O}(2.5\%)$ in the range of stop and gluino
masses allowed by the stronger bounds~(\ref{eq:15}). Both values are an
order of magnitude better than in the MSSM~\cite{Ghilencea:2012gz}.
(Points with $\Delta > 120$ have been rejected in our MCMC scans.)

\begin{figure}[ht!]
\begin{center}
\psfig{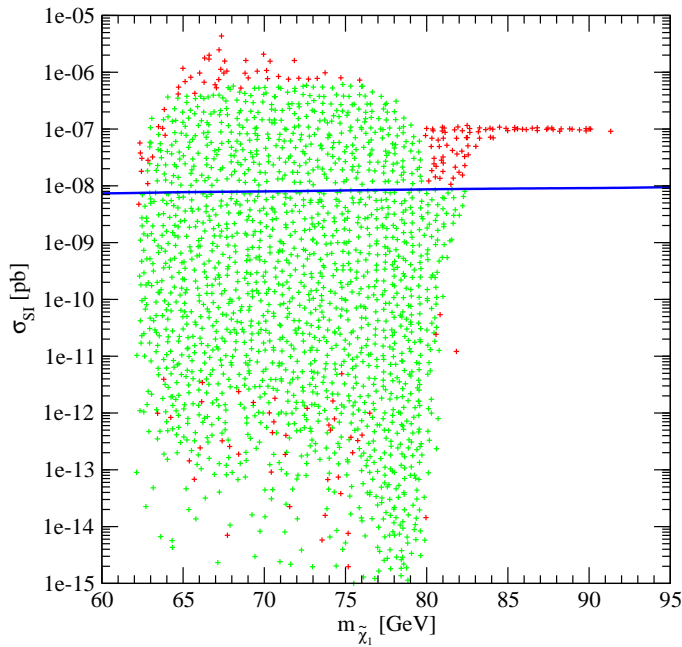}
\vspace*{3mm}
\caption{The spin-independent neutralino-proton scattering cross section
$\sigma^{si}(p)$ as a function of $M_{\chi^0_1}$. The blue line indicates
the bound from XENON100~\cite{Aprile:2011hi}, and we have added points
violating this bound (but respecting all the others). The color code is
as in Figs.~\ref{fig:4}.}
\label{fig:6}
\end{center}
\end{figure}

Turning to the neutralino sector we observe, as in the CP-even Higgs
sector, large mixing angles involving all 5 neutralinos of the NMSSM.
The lightest eigenstate $\chi^0_1$ (the LSP) is mostly higgsino-like, but
with sizeable bino, wino and singlino components and a mass in the
$60-90$~GeV range. Its direct detection cross section is reduced with
respect to pure higgsino-like neutralinos, and can well comply with the
constraints from XENON100. In Fig.~\ref{fig:6} we show the
spin-independent neutralino-proton scattering cross section
$\sigma^{si}(p)$ as a function of $M_{\chi^0_1}$.
We see that the stronger bounds on $M_{1/2}$ and $m_0$
in~(\ref{eq:15}) imply $60\ \text{GeV} \lsim M_{\chi^0_1} \lsim 85$~GeV,
similar to the range within the general NMSSM obtained in~\cite{Cao:2012fz}.
In particular, the plateau observed in Fig.~\ref{fig:6} for $80 \lsim M_{\chi^0_1}
\lsim 90$~GeV and $\sigma^{si}(p) \sim 10^{-7}$~pb, corresponding to small
values of $m_0, M_{1/2}$ and a mostly bino-like $\chi^0_1$, is excluded by
both the XENON100 and the strong LHC Susy contraints.
The spin-independent neutralino-proton scattering cross section
$\sigma^{si}(p)$ can vary over a wide range both above and below the
XENON100 limit~\cite{Aprile:2011hi} (which remains to be confirmed by
other experiments), but plenty of points would satisfy this constraint
and become observable in the future.

It may be interesting to know some of the properties of the Higgs and
sparticle sectors beyond the ones shown in the scatter plots above; to
this end we show two benchmark points in Table~1. The
point~(1) has $M_{H_1} \sim 100$~GeV, the point~(2) $M_{H_1} \sim
120$~GeV, and they differ in the values for $M_{1/2}$ and $m_0$.

\begin{table}
\begin{center}
\begin{tabular}{|c|c|c|c|} \hline
Point:                  & (1)      & (2)  \\\hline
Param. at $M_{GUT}$:                  &        &   \\\hline
$M_{1/2}$               & 600       & 525   \\\hline
$m_0$                   & 600       & 960   \\\hline
$A_0$                   & -1550       & -1140   \\\hline
$A_\lambda$             &  -625      &  -575  \\\hline
$A_\kappa$              &  -275      &  -360  \\\hline
$m_{H_u}$             & 1670     & 1880   \\\hline
$m_{H_d}$             & 445      & 757   \\\hline
$m_{S}$               & 885       & 1380   \\\hline
$\lambda$               & 0.96  &1.48    \\\hline
$\kappa$                & 0.73       & 1.08    \\\hline
$h_t$                   &0.83        &0.97    \\\hline
Param. at $M_{Susy}$:          &        &    \\\hline
$\lambda$               & 0.545       &  0.6  \\\hline
$\kappa$                & 0.253       & 0.321   \\\hline
$\tan\beta$             &  2.40      & 2.29   \\\hline
$\mu_{\text{eff}}$      &  120      & 122   \\\hline
Sparticle masses:       &        &    \\\hline
$m_{\tilde{g}}$       &  1390      &  1250  \\\hline
$m_{\tilde{q}}$ &   1320     &  1400  \\\hline
$m_{\tilde{t}_1}$     &   359     &  463  \\\hline
$m_{\tilde{b}_1}$    &   1001     &  1060  \\\hline
$m_{\tilde{\tau}_1}$       &   528     &  900  \\\hline
$M_{\chi^\pm_1}$ &    108    &  108  \\\hline
$M_{\chi^0_1}$            &    77    &  78  \\\hline
Components of $\chi^0_1$: &        &    \\\hline
$\tilde{B}$                    &   0.20     &  0.25  \\\hline
$\tilde{W}$                    &   -0.16     &  -0.20  \\\hline
$\tilde{H}_d$            &   0.48     &  0.52  \\\hline
$\tilde{H}_u$            &   -0.70     &  -0.70  \\\hline
$\tilde{S}$                &   0.46     &  0.37  \\\hline
$\Omega h^2$            &   0.10     & 0.10   \\\hline
$\sigma^{si}(p)\ [10^{-8}$~pb] &   1.00     & 0.13   \\\hline
$\Delta a_\mu\ [10^{-10}]$  & $0.93$ & $0.52$ \\\hline
\end{tabular}
\hspace*{10mm}
\begin{tabular}{|c|c|c|c|} \hline
Point:                  & (1)      & (2)  \\\hline
\hline
$M_{H_1}$  &   100     &  120  \\\hline
Components of $H_1$: &        &    \\\hline
$H_d$            &  0.39      &  0.50  \\\hline
$H_u$            &  0.34      & 0.74   \\\hline
$S$                    & 0.86       &  0.45  \\\hline
$R_1^{\gamma\gamma}(gg)$  &  0.01      &  0.32  \\\hline
$R_1^{\gamma\gamma}(\text{VBF})$  &   0.03     &  0.40  \\\hline
$R_1^{VV}(gg)$          &   0.03     &   0.34 \\\hline
$R_1^{\tau\tau}(\text{VBF})$  &   0.23     &  0.88  \\\hline
\hline
$M_{H_2}$  &   124     & 125   \\\hline
Components of $H_2$: &        &    \\\hline
$H_d$            &  0.26      & 0.04   \\\hline
$H_u$            &  0.85     &  -0.54  \\\hline
$S$                    &  -0.45      & 0.84   \\\hline
$R_2^{\gamma\gamma}(gg)$  &  1.54      &  1.77  \\\hline
$R_2^{\gamma\gamma}(\text{VBF})$  &   1.42     &  0.98  \\\hline
$R_2^{VV}(gg)$  &  1.22      &  1.01  \\\hline
$R_2^{\tau\tau}(\text{VBF})$  &  0.63      & 0.03   \\\hline\hline
$M_{H_3}$  &    329    &  305  \\\hline
Components of $H_3$: &        &    \\\hline
$H_d$            &  0.88      &  0.86  \\\hline
$H_u$            &  -0.40      & -0.40   \\\hline
$S$                    & -0.25       &  -0.30  \\\hline
$R_3^{\gamma\gamma}(gg)$  &  0.21      &  0.29  \\\hline
$R_3^{\gamma\gamma}(\text{VBF})$  & 0.0006       &  0.0007  \\\hline
$R_3^{VV}(gg)$  &  0.001      &  0.002  \\\hline
$R_3^{\tau\tau}(\text{VBF})$  &  0.04      &  0.03  \\\hline
\end{tabular}
\caption{Properties of two benchmark points corresponding to different
values of $M_{H_1}$.  All dimensionful parameters are given in
GeV. The components of $\chi^0_1$, as well as the components of $H_i$, are
defined such that their squares sum up to~1. The Susy contributions
$\Delta a_\mu$ to the muon anomalous magnetic moment are those given by
NMSSMTools without any theoretical errors.}
\end{center}
\end{table}

\section{Conclusions}

It has already been noted in~\cite{
Hall:2011aa,Ellwanger:2011aa,Arvanitaki:2011ck,King:2012is,Kang:2012tn,
Cao:2012fz} that the NMSSM can naturally accomodate Higgs bosons in the
$124-127$~GeV mass range. In addition, the NMSSM can explain excesses
in the $\gamma\,\gamma$ channel, as well as potential excesses at different
values of the Higgs mass (due to the extended Higgs sector). In the present
paper we have shown that these features persist in the constrained
NMSSM with non-universal Higgs sector, designated here as NUH-NMSSM.
The dominant deviation from full universality of all soft Susy breaking terms
at the GUT scale originates from the need to have $m_{H_u} > m_0$.

The following properties of the Higgs sector are peculiar:
\begin{itemize}
\item the signal rate in the $\gamma\,\gamma$ channel can be 2.8 as
large as the one of a SM-like Higgs boson, provided the mass of the
lighter CP-even state $H_1$ is in the $115 - 123$~GeV range;
\item requiring a visible signal rate in the $b\bar{b}$ channel of 0.9
times the SM value allows for a signal rate in the $\gamma\,\gamma$
channel about 1.6 as large as the one of a SM-like Higgs boson; \item
the lighter CP-even state $H_1$ could explain a mild excess of events
around $95-100$~GeV observed at LEP, or a second visible Higgs boson
below $\sim 123$~GeV.
\end{itemize}

In the sparticle sector, the assumption of universality at the GUT scale
leads to the following features:
\begin{itemize}
\item the mass of the lightest stop can be as small as 105~GeV,
complying with present constraints for $M_{\tilde{g}} \gsim 640$~GeV;
\item the fine-tuning with respect to parameters at the GUT scale
remains modest, an order of magnitude below the one required in the
MSSM;
\item the eigenstates in the neutralino sector are strongly mixed, and
the lightest neutralino can have a relic density in agreement with WMAP
constraints. Its direct detection cross section can be above or below
present XENON100 bounds; most of the points below these bounds should be
observable in the near future.
\end{itemize}

Given the large values of the NMSSM-specific coupling $\lambda$, all
scenarios presented here differ strongly from the MSSM (also by the low
value of $\tan\beta$). The fact that all 3 Yukawa couplings $\lambda$,
$\kappa$ and $h_t$ are of ${\cal O}(1)$ at the GUT scale may hint at
some strong dynamics present at that scale. It is possible that the
deviation from full universality of soft Susy breaking terms at the GUT
scale remains confined to $m_{H_u} > m_0$; such possibilities require
further studies.

Of course, first of all the present evidence for a Higgs boson in the
$124 - 127$~GeV mass range
should be confirmed by more data;
then possible evidence for non-SM properties of the Higgs sector
like an enhanced cross section in the diphoton channel
will show whether the scenarios presented here are realistic.

\section*{Acknowledgements}

U. E. acknowledges support from the French ANR LFV-CPV-LHC.
C. H. acknowledges support from the French ANRJ TPADMS.


\begin{thebibliography}{99}

\bibitem{:2012si}
  G.~Aad {\it et al.}  [ATLAS Collaboration],
  ``Combined search for the Standard Model Higgs boson using up to 4.9
  fb-1 of pp collision data at sqrt(s) = 7 TeV with the ATLAS detector
  at the LHC,'' arXiv:1202.1408 [hep-ex].

\bibitem{Chatrchyan:2012tx}
  S.~Chatrchyan {\it et al.}  [CMS Collaboration],
  ``Combined results of searches for the standard model Higgs boson in
  pp collisions at sqrt(s) = 7 TeV,'' arXiv:1202.1488 [hep-ex].

\bibitem{:2012sk}
  [ATLAS Collaboration],
  ``Search for the Standard Model Higgs boson in the diphoton decay channel with 4.9 fb-1 of pp collisions at sqrt(s)=7 TeV with ATLAS,''
  arXiv:1202.1414 [hep-ex].

\bibitem{Chatrchyan:2012tw}
  S.~Chatrchyan {\it et al.}  [CMS Collaboration],
  ``Search for the standard model Higgs boson decaying into two photons in pp collisions at sqrt(s)=7 TeV,''
  arXiv:1202.1487 [hep-ex].
  
\bibitem{ATLAS-CONF-2012-019} ATLAS Collaboration, ``An update to the
combined search for the Standard Model Higgs boson with the ATLAS
detector at the LHC using up to 4.9~$fb^{-1}$ of $pp$ collision data at
$\sqrt{s}=7$~TeV, ATLAS-CONF-2012-019.

\bibitem{CMS-PAS-HIG-12-001} CMS Collaboration, ``A search using
multivaraite techniques for a standard model Higgs boson decaying into
two photons'', CMS-PAS-HIG-12-001.

\bibitem{Hall:2011aa}
  L.~J.~Hall, D.~Pinner and J.~T.~Ruderman,
  ``A Natural SUSY Higgs Near 126 GeV,''
  arXiv:1112.2703 [hep-ph].

\bibitem{Baer:2011ab}
  H.~Baer, V.~Barger and A.~Mustafayev,
  ``Implications of a 125 GeV Higgs scalar for LHC SUSY and neutralino dark matter searches,''
  arXiv:1112.3017 [hep-ph].

\bibitem{Feng:2011aa}
  J.~L.~Feng, K.~T.~Matchev and D.~Sanford,
  ``Focus Point Supersymmetry Redux,''
  arXiv:1112.3021 [hep-ph].

\bibitem{Heinemeyer:2011aa}
  S.~Heinemeyer, O.~Stal and G.~Weiglein,
  ``Interpreting the LHC Higgs Search Results in the MSSM,''
  arXiv:1112.3026 [hep-ph].
 
\bibitem{Arbey:2011ab}
  A.~Arbey, M.~Battaglia, A.~Djouadi, F.~Mahmoudi and J.~Quevillon,
  Phys.\ Lett.\ B {\bf 708} (2012) 162
  [arXiv:1112.3028 [hep-ph]].

\bibitem{Arbey:2011aa}
  A.~Arbey, M.~Battaglia and F.~Mahmoudi,
  Eur.\ Phys.\ J.\ C {\bf 72} (2012) 1906
  [arXiv:1112.3032 [hep-ph]].

\bibitem{Draper:2011aa}
  P.~Draper, P.~Meade, M.~Reece and D.~Shih,
  ``Implications of a 125 GeV Higgs for the MSSM and Low-Scale SUSY Breaking,''
  arXiv:1112.3068 [hep-ph].
  
\bibitem{Moroi:2011aa}
  T.~Moroi, R.~Sato and T.~T.~Yanagida,
  Phys.\ Lett.\ B {\bf 709} (2012) 218
  [arXiv:1112.3142 [hep-ph]].

\bibitem{Carena:2011aa}
  M.~Carena, S.~Gori, N.~R.~Shah and C.~E.~M.~Wagner,
  ``A 125 GeV SM-like Higgs in the MSSM and the $\gamma \gamma$ rate,''
  arXiv:1112.3336 [hep-ph].

\bibitem{Ellwanger:2011aa}
  U.~Ellwanger,
  ``A Higgs boson near 125 GeV with enhanced diphoton signal in the NMSSM,''
  arXiv:1112.3548 [hep-ph].

\bibitem{Buchmueller:2011ab}
  O.~Buchmueller, R.~Cavanaugh, A.~De Roeck, M.~J.~Dolan, J.~R.~Ellis, H.~Flacher, S.~Heinemeyer and G.~Isidori {\it et al.},
  ``Higgs and Supersymmetry,''
  arXiv:1112.3564 [hep-ph].

\bibitem{Akula:2011aa}
  S.~Akula, B.~Altunkaynak, D.~Feldman, P.~Nath and G.~Peim,
  ``Higgs Boson Mass Predictions in SUGRA Unification, Recent LHC-7
  Results, and Dark Matter,'' arXiv:1112.3645 [hep-ph].

\bibitem{Kadastik:2011aa}
  M.~Kadastik, K.~Kannike, A.~Racioppi and M.~Raidal,
  ``Implications of 125 GeV Higgs boson on scalar dark matter and on the
  CMSSM phenomenology,'' arXiv:1112.3647 [hep-ph].
 
\bibitem{Cao:2011sn}
  J.~Cao, Z.~Heng, D.~Li and J.~M.~Yang, ``Current experimental
  constraints on the lightest Higgs boson mass in the constrained
  MSSM,'' arXiv:1112.4391 [hep-ph].

\bibitem{Arvanitaki:2011ck}
  A.~Arvanitaki and G.~Villadoro,
  ``A Non Standard Model Higgs at the LHC as a Sign of Naturalness,''
  arXiv:1112.4835 [hep-ph].

\bibitem{Gozdz:2012xx}
  M.~Gozdz,
  ``Lightest Higgs boson masses in the R-parity violating supersymmetry,''
  arXiv:1201.0875 [hep-ph].
  
\bibitem{Gunion:2012zd}
  J.~F.~Gunion, Y.~Jiang and S.~Kraml,
  ``The Constrained NMSSM and Higgs near 125 GeV,''
  arXiv:1201.0982 [hep-ph].

\bibitem{FileviezPerez:2012iw}
  P.~Fileviez Perez,
  ``SUSY Spectrum and the Higgs Mass in the BLMSSM,''\\
  arXiv:1201.1501 [hep-ph].

\bibitem{Karagiannakis:2012vk}
  N.~Karagiannakis, G.~Lazarides and C.~Pallis,
  ``Dark Matter and Higgs Mass in the CMSSM with Yukawa Quasi-Unification,''
  arXiv:1201.2111 [hep-ph].

\bibitem{King:2012is}
  S.~F.~King, M.~Muhlleitner and R.~Nevzorov,
  ``NMSSM Higgs Benchmarks Near 125 GeV,''
  arXiv:1201.2671 [hep-ph].
  
\bibitem{Kang:2012tn}
  Z.~Kang, J.~Li and T.~Li,
  ``On the Naturalness of the (N)MSSM,''
  arXiv:1201.5305 [hep-ph].

\bibitem{Chang:2012gp}
  C.~-F.~Chang, K.~Cheung, Y.~-C.~Lin and T.~-C.~Yuan,
  ``Mimicking the Standard Model Higgs Boson in UMSSM,''
  arXiv:1202.0054 [hep-ph].
  
\bibitem{Aparicio:2012iw}
  L.~Aparicio, D.~G.~Cerdeno and L.~E.~Ibanez,
  ``A 119-125 GeV Higgs from a string derived slice of the CMSSM,''
  arXiv:1202.0822 [hep-ph].
  
\bibitem{Roszkowski:2012uf}
  L.~Roszkowski, E.~M.~Sessolo and Y.~-L.~S.~Tsai,
  ``Bayesian Implications of Current LHC Supersymmetry and Dark Matter Detection Searches for the Constrained MSSM,''
  arXiv:1202.1503 [hep-ph].
  
\bibitem{Ellis:2012aa}
  J.~Ellis and K.~A.~Olive,
  ``Revisiting the Higgs Mass and Dark Matter in the CMSSM,''
  arXiv:1202.3262 [hep-ph].

\bibitem{Baer:2012uy}
  H.~Baer, V.~Barger and A.~Mustafayev,
  ``Neutralino dark matter in mSUGRA/CMSSM with a 125 GeV light Higgs scalar,''
  arXiv:1202.4038 [hep-ph].

\bibitem{Desai:2012qy}
  N.~Desai, B.~Mukhopadhyaya and S.~Niyogi,
  ``Constraints on invisible Higgs decay in MSSM in the light of diphoton rates from the LHC,''
  arXiv:1202.5190 [hep-ph].
  
\bibitem{Cao:2012fz}
  J.~Cao, Z.~Heng, J.~M.~Yang, Y.~Zhang and J.~Zhu,
  ``A SM-like Higgs near 125 GeV in low energy SUSY: a comparative study for MSSM and NMSSM,''
  arXiv:1202.5821 [hep-ph].

\bibitem{Maiani:2012ij}
  L.~Maiani, A.~D.~Polosa and V.~Riquer,
  ``Probing Minimal Supersymmetry at the LHC with the Higgs Boson Masses,''
  arXiv:1202.5998 [hep-ph].
  
\bibitem{Cheng:2012np}
  T.~Cheng, J.~Li, T.~Li, D.~V.~Nanopoulos and C.~Tong,
  ``Electroweak Supersymmetry around the Electroweak Scale,''
  arXiv:1202.6088 [hep-ph].

\bibitem{Christensen:2012ei}
  N.~Christensen, T.~Han and S.~Su,
  ``MSSM Higgs Bosons at The LHC,''
  arXiv:1203.3207 [hep-ph].

\bibitem{Vasquez:2012hn}
  D.~A.~Vasquez, G.~Belanger, C.~Boehm, J.~Da Silva, P.~Richardson and C.~Wymant,
  ``The 125 GeV Higgs in the NMSSM in light of LHC results and astrophysics
  constraints,''
  arXiv:1203.3446 [hep-ph].

\bibitem{Maniatis:2009re}
  M.~Maniatis,
  Int.\ J.\ Mod.\ Phys.\  A {\bf 25} (2010) 3505
  [arXiv:0906.0777 [hep-ph]].

\bibitem{Ellwanger:2009dp}
  U.~Ellwanger, C.~Hugonie and A.~M.~Teixeira,
  Phys.\ Rept.\  {\bf 496} (2010) 1\newline
  [arXiv:0910.1785 [hep-ph]].
:

\bibitem{Djouadi:2008yj}
  A.~Djouadi, U.~Ellwanger and A.~M.~Teixeira,
  Phys.\ Rev.\ Lett.\  {\bf 101} (2008) 101802
  [arXiv:0803.0253 [hep-ph]].

\bibitem{Djouadi:2008uj}
  A.~Djouadi, U.~Ellwanger and A.~M.~Teixeira,
  JHEP {\bf 0904} (2009) 031
  [arXiv:0811.2699 [hep-ph]].

\bibitem{Komatsu:2010fb}
  E.~Komatsu {\it et al.}  [WMAP Collaboration],
  Astrophys.\ J.\ Suppl.\  {\bf 192} (2011) 18
  [arXiv:1001.4538 [astro-ph.CO]].

\bibitem{Chatrchyan:2011zy}
  S.~Chatrchyan {\it et al.}  [CMS Collaboration],
  Phys.\ Rev.\ Lett.\  {\bf 107} (2011) 221804
  [arXiv:1109.2352 [hep-ex]].
  
\bibitem{Aad:2011ib}
  G.~Aad {\it et al.}  [ATLAS Collaboration],
  ``Search for squarks and gluinos using final states with jets and
  missing transverse momentum with the ATLAS detector in $\sqrt(s)$ =
  7~TeV proton-proton collisions,'' arXiv:1109.6572 [hep-ex].

\bibitem{CMS-PAS-SUS-12-005}
  CMS collaboration, ``Search for supersymmetry with the razor variables
  at CMS'', CMS-PAS-SUS-12-005

\bibitem{Aprile:2011hi}
  E.~Aprile {\it et al.}  [XENON100 Collaboration],
  Phys.\ Rev.\ Lett.\  {\bf 107} (2011) 131302
  [arXiv:1104.2549 [astro-ph.CO]].

\bibitem{Schael:2006cr}
  S.~Schael {\it et al.} [ALEPH and DELPHI and L3 and OPAL 
  Collaborations and LEP Working Group for Higgs Boson Searches],
  Eur.\ Phys.\ J.\  C {\bf 47} (2006) 547
  [arXiv:hep-ex/0602042].

\bibitem{Das:2012rr}
  D.~Das, U.~Ellwanger and A.~M.~Teixeira,
  ``Modified Signals for Supersymmetry in the NMSSM with a Singlino-like
  LSP,'' arXiv:1202.5244 [hep-ph].

\bibitem{ATLAS-stop} ATLAS collaboration, ``Search for supersymmetry in
pp collisions at $\sqrt{s}$ = 7~TeV in final states with missing transverse
momentum and b-jets with the ATLAS detector'', ATLAS-CONF-2012-003.

\bibitem{Tevatron-stop} D0 and CDF collaborations, ``Search for scalar
top and bottom quarks at the Tevatron'', FERMILAB-CONF-09-108-E

\bibitem{Bennett:2006fi}
  G.~W.~Bennett {\it et al.}  [Muon G-2 Collaboration],
  Phys.\ Rev.\ D {\bf 73} (2006) 072003
  [hep-ex/0602035].
 
\bibitem{LEPSUSY}
LEPSUSYWG, ALEPH, DELPHI, L3 and OPAL experiments,\newline
 notes LEPSUSYWG/04-01.1 and 04-02.1
(http://lepsusy.web.cern.ch/lepsusy/).

\bibitem{Abazov:2006wb} 
  V.~M.~Abazov {\it et al.}  [D0 Collaboration],
  Phys.\ Lett.\ B {\bf 645}, 119 (2007)
  [hep-ex/0611003].

\bibitem{Wang:2009zzf}
  S.~M.~Wang  [CDF and D0 Collaborations],
  AIP Conf.\ Proc.\  {\bf 1078} (2009) 259.

\bibitem{Ellwanger:2006rn}
  U.~Ellwanger and C.~Hugonie,
  Comput.\ Phys.\ Commun.\  {\bf 177} (2007) 399
  [hep-ph/0612134].

\bibitem{Ellwanger:2004xm}
  U.~Ellwanger, J.~F.~Gunion and C.~Hugonie,
  JHEP {\bf 0502} (2005) 066
  [arXiv:hep-ph/0406215].

\bibitem{Ellwanger:2005dv}
  U.~Ellwanger and C.~Hugonie,
  Comput.\ Phys.\ Commun.\  {\bf 175} (2006) 290
  [arXiv:hep-ph/0508022].

\bibitem{Degrassi:2009yq}
  G.~Degrassi and P.~Slavich,
  Nucl.\ Phys.\  B {\bf 825} (2010) 119
  [arXiv:0907.4682 [hep-ph]].

\bibitem{Chatrchyan:2012vp}
  S.~Chatrchyan {\it et al.}  [CMS Collaboration],
  ``Search for neutral Higgs bosons decaying to tau pairs in pp collisions at sqrt(s)=7 TeV,''
  arXiv:1202.4083 [hep-ex].

\bibitem{Belanger:2005kh}
  G.~Belanger, F.~Boudjema, C.~Hugonie, A.~Pukhov and A.~Semenov,
  JCAP {\bf 0509} (2005) 001
  [arXiv:hep-ph/0505142].

\bibitem{Belanger:2006is}
  G.~Belanger, F.~Boudjema, A.~Pukhov and A.~Semenov,
  Comput.\ Phys.\ Commun.\  {\bf 176}, 367 (2007)
  [arXiv:hep-ph/0607059].

\bibitem{Belanger:2008sj}
  G.~Belanger, F.~Boudjema, A.~Pukhov and A.~Semenov,
  Comput.\ Phys.\ Commun.\  {\bf 180} (2009) 747
  [arXiv:0803.2360 [hep-ph]].

\bibitem{Ellwanger:2010nf}
  U.~Ellwanger,
  Phys.\ Lett.\ B {\bf 698} (2011) 293
  [arXiv:1012.1201 [hep-ph]].

\bibitem{Tevatron-higgs} CDF and D0 collaborations,
  ``Combined CDF and D0 Searches for Standard Model Higgs Boson
  Production'', FERMILAB-CONF-12-065-E, CDF Note 10806, D0 Note 63032

\bibitem{Dermisek:2007ah}
  R.~Dermisek and J.~F.~Gunion,
  Phys.\ Rev.\ D {\bf 77} (2008) 015013
  [arXiv:0709.2269 [hep-ph]].

\bibitem{Ellwanger:2011sk}
  U.~Ellwanger,
  Eur.\ Phys.\ J.\ C {\bf 71} (2011) 1782
  [arXiv:1108.0157 [hep-ph]].

\bibitem{Ellwanger:2011mu}
  U.~Ellwanger, G.~Espitalier-Noel and C.~Hugonie,
  JHEP {\bf 09} (2011) 105\newline  [arXiv:1107.2472 [hep-ph]].

\bibitem{Ghilencea:2012gz}
  D.~M.~Ghilencea, H.~M.~Lee and M.~Park,
  ``Tuning supersymmetric models at the LHC: A comparative analysis at two-loop level,''
  arXiv:1203.0569 [hep-ph].

\end{thebibliography}
\end{document}